\documentclass[12pt, a4paper]{elsarticle}
\makeatletter
\def\ps@pprintTitle{%
 \let\@oddhead\@empty
 \let\@evenhead\@empty
 \let\@oddfoot\@empty
 \let\@evenfoot\@oddfoot
}
\makeatother
\usepackage{amsmath}
\usepackage{amssymb}
\usepackage{subfigure}
\usepackage{xspace}
\usepackage{xcolor}
\usepackage{hyperref}
\usepackage{lineno}
\usepackage{float}
\usepackage{listings}
\restylefloat{table}
\newcommand{\BLUE     }{{\color{blue}{\sc BLUE}}}
\newcommand{\chiq   }{\ensuremath{\chi^2}\xspace}
\newcommand{\xo     }{\ensuremath{x_\mathrm{1}}\xspace}
\newcommand{\xt     }{\ensuremath{x_\mathrm{2}}\xspace}
\newcommand{\Xo     }{\ensuremath{X_\mathrm{1}}\xspace}
\newcommand{\Xt     }{\ensuremath{X_\mathrm{2}}\xspace}
\newcommand{\xT     }{\ensuremath{x_\mathrm{T}}\xspace}
\newcommand{\si     }{\ensuremath{\sigma_{i}}\xspace}

\newcommand{\so     }{\ensuremath{\sigma_{1}}\xspace}
\newcommand{\soq    }{\ensuremath{\sigma_{1}^{2}}\xspace}
\newcommand{\st     }{\ensuremath{\sigma_{2}}\xspace}
\newcommand{\stq    }{\ensuremath{\sigma_{2}^{2}}\xspace}

\newcommand{\sdq    }{\ensuremath{\sigma_{\Delta}^{2}}\xspace}
\newcommand{\sx     }{\ensuremath{\sigma_{x}}\xspace}
\newcommand{\sxq    }{\ensuremath{\sigma_{x}^{2}}\xspace}
\newcommand{\rhof   }{\ensuremath{\rho}\xspace}
\newcommand{\rhoij  }{\ensuremath{\rho_{ij}}\xspace}
\newcommand{\rhoijk }{\ensuremath{\rho_{ijk}}\xspace}
\newcommand{\rhono  }{\ensuremath{\rho_{01}}\xspace}
\newcommand{\rhont  }{\ensuremath{\rho_{02}}\xspace}
\newcommand{\rhoot  }{\ensuremath{\rho_{12}}\xspace}
\newcommand{\bet    }{\ensuremath{\beta}\xspace}
\newcommand{\M      }{\ensuremath{M}\xspace}
\newcommand{\Mz     }{\ensuremath{M_\mathrm{0}}\xspace}
\newcommand{\Mo     }{\ensuremath{M_\mathrm{1}}\xspace}
\newcommand{\Mt     }{\ensuremath{M_\mathrm{2}}\xspace}
\newcommand{\as     }{\ensuremath{\alpha_{s}}\xspace}
\newcommand{\sigJES }{\ensuremath{\sigma_{\mathrm{JES, l}}}\xspace}
\newcommand{\Sys    }[1]{\ensuremath{\mathrm{Syst}_{#1}}\xspace}
\newcommand{\Key    }[1]{\ensuremath{\mathtt{\underline{#1}:}}\xspace}
\begin{document}
\begin{frontmatter}
\title{\BLUE: combining correlated estimates of physics observables within ROOT
  using the Best Linear Unbiased Estimate method}
\author{Richard Nisius}
\address{Max-Planck-Institut f\"ur Physik (Werner-Heisenberg-Institut),
  F\"ohringer Ring 6, D-80805 M\"unchen, Germany}
%
%
\begin{abstract}
 This software performs the combination of $m$ correlated estimates of $n$
 physics observables~($m\ge n$) using the Best Linear Unbiased Estimate (\BLUE)
 method.
 It is implemented as a \verb!C++! class, to be used within the ROOT analysis
 package.
 It features easy disabling of specific estimates or uncertainty sources, the
 investigation of different correlation assumptions, and allows performing
 combinations according to the importance of the estimates.
 This enables systematic investigations of the combination on details of the
 measurements from within the software, without touching the input.
\end{abstract}
%
%
\end{frontmatter}
%
\begin{table}[H]
\footnotesize
\begin{tabular}{p{6.0cm}p{6.0cm}}
\textbf{Code metadata} & \textbf{} \\
\hline
 Current code version & 2.4.0 \\
 Permanent link to code/repository used &
 {\href{https://github.com/ElsevierSoftwareX/SOFTX\_2020\_16}
       {https://github.com/ElsevierSoftwareX/SOFTX\_2020\_16}} \\
 for this code version &
 {\href{https://blue.hepforge.org}{https://blue.hepforge.org}} \\
 Code Ocean compute capsule & \\
 Legal Code License & LGPL \\
 Code versioning system used & none \\
 Software code languages, tools, and services used & \verb!C++! \\
 Compilation requirements, operating environments \& dependencies &
 depends on
 {\href{https://root.cern.ch}{https://root.cern.ch}} \\
 If available Link to developer documentation/manual &
 {\href{https://blue.hepforge.org}{https://blue.hepforge.org}} \\
 Support email for questions & Richard.Nisius@mpp.mpg.de \\\hline
\end{tabular}
\end{table}
%
%
\section{Motivation and significance}
\label{sec:moti}
 The combination of a number of correlated estimates for a single observable is
 discussed in Ref.~\cite{LYO-1988}. Here, the term estimate denotes a particular
 outcome (measurement) of an experiment based on an experimental estimator~(an
 algorithm for a measurement) of the observable, which follows a probability
 density function~(PDF). The particular estimate obtained by the experiment may
 be a likely or unlikely outcome for that PDF. Repeating the measurement
 numerous times under identical conditions, the estimates will follow the
 underlying PDF of the estimator.

 The metadata for the \BLUE\ software are listed in the Code metadata table. The
 software uses the Gaussian approximation for the uncertainties and performs a
 \chiq\ minimization to obtain the combined value. In Ref.~\cite{LYO-1988}, this
 minimization is expressed using the mathematically equivalent \BLUE\ ansatz.
 Provided the estimators are unbiased, when applying this formalism the
 {\bf Best Linear Unbiased Estimate}
 of the observable is obtained with the following meaning:
 {\bf Best:} the combined result for the observable obtained this way has the
 smallest variance; {\bf Linear:} the result is constructed as a linear
 combination of the individual estimates; {\bf Unbiased Estimate:} when the
 procedure is repeated for a large number of cases consistent with the
 underlying multi-dimensional PDF, the mean of all combined results equals the
 true value of the observable.
 The formulas for more than one observable~\cite{VAL-0301} are implemented in
 the \BLUE\ software, which is programmed as a separate class of the ROOT
 analysis framework~\cite{BRU-9701}.

\begin{figure*}[tbp!]
\centering
\subfigure[\bet as a function of \rhof]{
   \label{fig:beta}\includegraphics[width=0.48\textwidth]{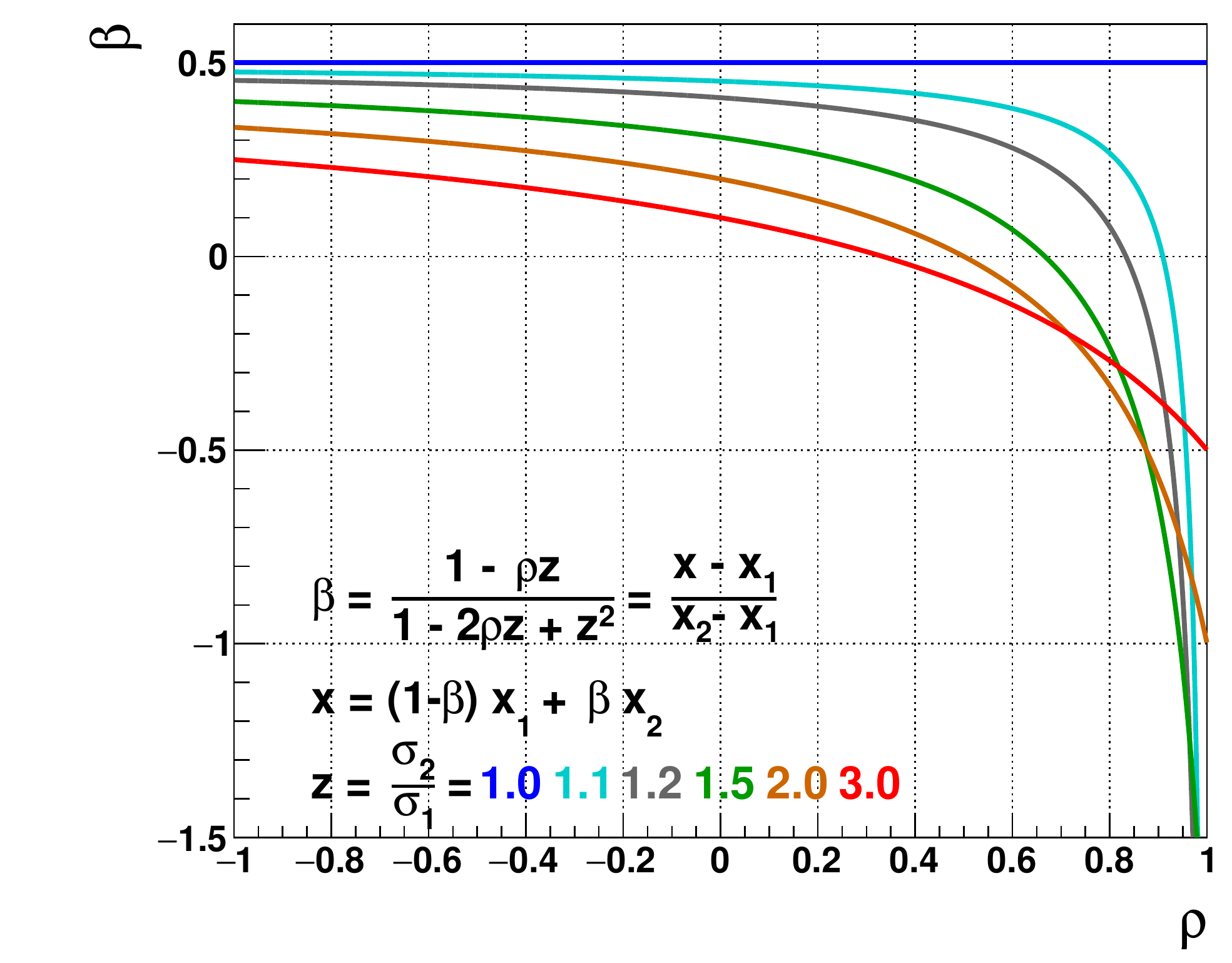}}
\subfigure[\sx/\so as a function of \rhof]{
   \label{fig:sigx}\includegraphics[width=0.48\textwidth]{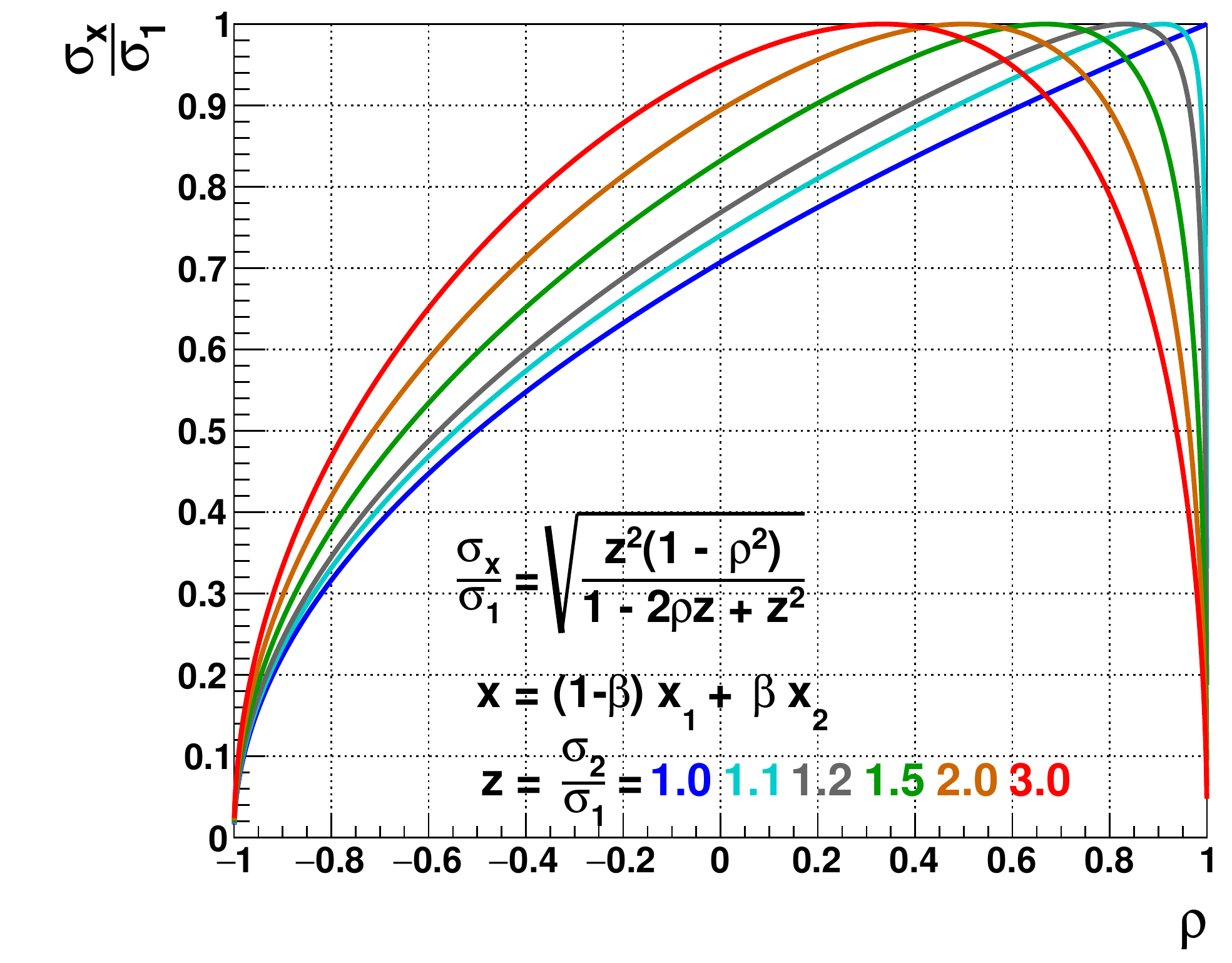}}
\caption{The variables \bet\ (a) and \sx/\so\ (b) shown as functions of
  \rhof\ for a number of $z$ values.
  \label{fig:para}
}
\end{figure*}

 The easiest case of two correlated estimates of the same observable is briefly
 illustrated here. Already for this case, the main features of the combination
 can easily be understood. For further information and the derivation of the
 formulas the reader is refereed to Ref.~\cite{NIS-1401}.
 Let \xo\ and \xt\ with variances \soq\ and \stq\ be two estimates from two
 unbiased estimators \Xo\ and \Xt\ of the true value \xT\ of the observable and
 \rhof\ the total correlation of the two estimators.
 Without loss of generality it is assumed that the estimate \xo\ stems from an
 estimator \Xo\ of \xT\ that is at least as precise as the estimator
 \Xt\ yielding the estimate \xt, such that $z\equiv \st/\so\geq 1$.
 In this situation the \BLUE\ $x$ of \xT\ is:
%
\begin{eqnarray*}
   x &=\,& (1-\bet)\,\xo + \bet\,\xt\, \nonumber
\end{eqnarray*}
%
 where \bet\ is the weight of the less precise estimate and the sum of weights
 is unity by construction. The variable $x$ is the combined result and \sxq\ is
 its variance, i.e.~the uncertainty assigned to the combined value is \sx.
 For a number of $z$ values the two quantities \bet\ and \sx/\so\ as functions
 of \rhof\ are shown in Fig.~\ref{fig:para}. Their functional forms are also
 written in the figures.
 The functions are valid for $-1 \leq \rhof \leq 1$ and $z \geq 1$, except for
 $\rhof = z = 1$.
%

 Fig.~\ref{fig:para} displays the strong dependence of the uncertainty in the
 combined value on $z$ and \rhof. For the special situation of $\rhof = 1/z$ the
 uncertainty \sx\ equals \so, i.e.~the precision in the observable is not
 improved by adding the second measurement.
 For $\rhof=\pm1$ the uncertainty in the combined result vanishes, i.e.~$\sx=0$,
 a mere consequence of the conditional probability for \Xt\ given the measured
 value of \xo, see Ref.~\cite{NIS-1401} for details.
 It is worth noticing that in most regions of the (\rhof, $z$)-plane the
 sensitivity of \sx/\so\ to \rhof\ is stronger than to $z$. This means, reducing
 the correlation of the estimates in most cases gives a larger improvement in
 precision in the combined value than reducing $z$.
%
\section{Software description}
\label{sec:desc}
%
\begin{figure*}[tbp!]
\centering
\includegraphics[width=0.90\textwidth]{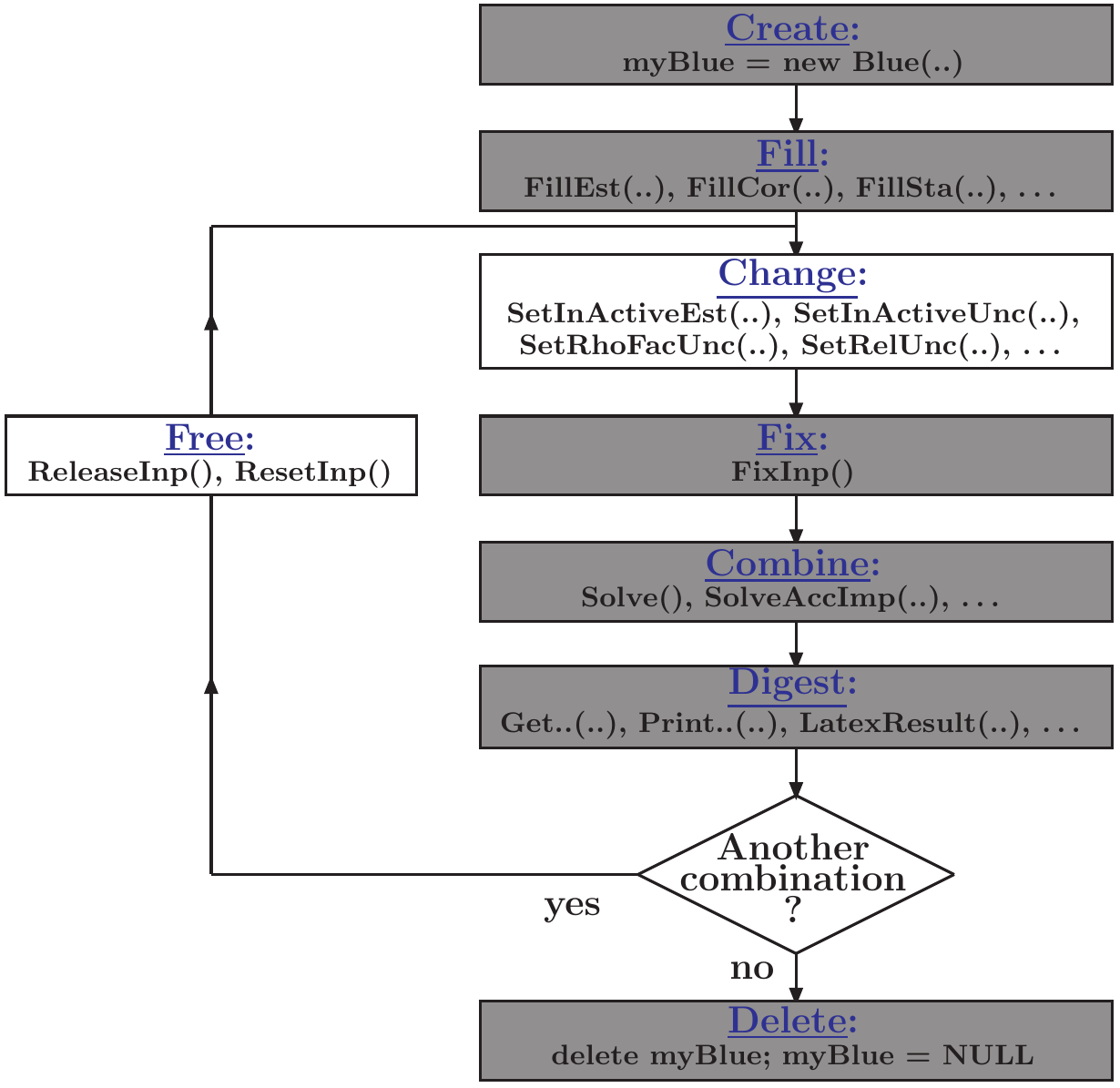}
\caption{Flowchart of the \BLUE\ software. The gray boxes display the steps to
  be performed for a single combination with unchanged inputs. The white boxes
  display the optional steps for further combinations with changed inputs and/or
  different solving methods.
  In each rectangular box, the first line indicates the performed action, while
  the following line(s) list the functions to use.
\label{fig:work}
}
\end{figure*}
%
To use the \BLUE\ combination software, a working installation of the ROOT
 package~\cite{BRU-9701} is needed.
 The software allows repeated combinations interleaved with an arbitrary number
 of changes to various aspects of the input, achieved using {\tt Set\ldots()}
 functions.
 In addition, it gives access to many intermediate quantities used in the
 combination via {\tt Get\ldots()} and {\tt Print\ldots()} functions. Only at
 well-defined steps of the computation, will the returned quantities be
 meaningful.
 To enable the above flexibility, the \BLUE\ object has several states like {\tt
   IsFilledInp}, {\tt IsFixed} or {\tt IsSolved} that can be false or true. The
 various functions are only enabled, if the object has the appropriate state.

 The mandatory inputs to the software are the measured values, their
 uncertainties for the various statistical and systematic effects relevant to
 the measurements, and the estimator correlations for each of those
 uncertainties.
 Those are inserted into the \BLUE\ object by means of the filling functions
 {\tt FillEst()} and {\tt FillCor()}.
 The end of filling the mandatory inputs is automatically recognized by the
 software.
 At this point, the initial input is saved, auxiliary information is calculated,
 {\tt IsFilledInp} is set to true and all filling functions are disabled.
 Consequently, the optional additional input like names, print formats and a
 logo, all have to be filled before the end of filling the mandatory input is
 reached.

 The most important object states can be controlled by function calls.
 For example, a call to {\tt FixInp()} initiates the calculation of various
 quantities and sets {\tt IsFixed} to true, thereby enabling the use of a number
 of {\tt Get\ldots()} and {\tt Print\ldots()} functions.
 A subsequent call to any of the solving methods {\tt Solve\ldots()} performs
 the desired combination, and finally sets {\tt IsSolved} to true.
 In this scheme, solving is only possible after a call to {\tt FixInp()}, and
 changing the inputs only after a call to {\tt ReleaseInp()} or {\tt
   ResetInp()}.
 Internally, the combination is always performed on a temporary copy of the
 initial input, which can be successively altered using the sequence of calling
 {\tt ReleaseInp()}, {\tt Set\ldots()} and {\tt FixInp()}.
 Since the initial input was saved as described above, a call to {\tt
   ResetInp()} allows to return to this situation.
%
\subsection{Combination workflow}
\label{sec:arch}
 The flowchart of the software is shown in Fig.~\ref{fig:work}.
 After calling the constructor of the class~(\Key{Create}), thereby defining the
 number of estimates, uncertainties and observables, the measurements together
 with their uncertainties and correlations are passed to the
 software~(\Key{Fill}).
 If wanted, the inputs can be adapted~(\Key{Change}).
 Combinations are performed following the loop of fixing~(\Key{Fix}),
 combining~(\Key{Combine}) and evaluating the result~(\Key{Digest}).
 Before changing the inputs for the next combination, they have to be
 freed~(\Key{Free}). For this step, two options are available, ReleaseInp()
 means the change proceeds from the status at the last fix, while {\tt
   ResetInp()} reverts to the original inputs.
 If no further combinations are wanted, the object is deleted~(\Key{Delete}).
%
\subsection{Software Functionalities}
\label{sec:func}
 One main quality of this software is the built-in flexibility for easy and
 thorough investigations of the impact of details of the input measurements and
 their correlation.
 With single function calls estimates or uncertainty sources can be removed from
 the combination, different uncertainty models (e.g.~absolute or relative
 uncertainties) and correlation assumptions can be investigated.
 Another strength is the large number of different solving methods implemented,
 ranging from only using measurements with positive weights in the combination
 to a successive combination method in which the input measurements are included
 one at a time according to their importance, allowing an in-depth investigation
 of their impact on the combination.
%
\section{Illustrative Example}
\label{sec:exam}
%
\begin{table}[t!]
\footnotesize
\begin{center}
\begin{tabular}{lrrrrrrr} \hline
 \multicolumn{1}{c}{}             &\multicolumn{3}{c}{Estimates} 
&\multicolumn{3}{c}{Correlations} &\multicolumn{ 1}{c}{Result} \\ \hline
         &             \Mz &             \Mo &             \Mt & \rhono  & \rhont  & \rhoot &      \M \\
         &           [GeV] &           [GeV] &           [GeV] &         &         &         &  [GeV]\\\hline
   Value &          174.86 &          172.63 &          173.25 &         &         &         & 173.92\\ 
    Stat &            0.35 &            0.54 &            0.24 &         &         &         &   0.20\\ 
 \Sys{1} & 0.26 $\pm$ 0.06 & 0.66 $\pm$ 0.04 & 0.43 $\pm$ 0.06 & $+1.00$ & $+1.00$ & $+1.00$ &   0.36\\
 \Sys{2} & 0.09 $\pm$ 0.05 & 0.64 $\pm$ 0.06 & 0.23 $\pm$ 0.08 & $+1.00$ & $+1.00$ & $+1.00$ &   0.17\\
 \Sys{3} & 0.12 $\pm$ 0.14 & 0.47 $\pm$ 0.09 & 0.23 $\pm$ 0.11 & $-1.00$ & $-1.00$ & $-1.00$ &   0.09\\
 \Sys{4} & 0.18 $\pm$ 0.08 & 0.24 $\pm$ 0.05 & 0.10 $\pm$ 0.08 & $-1.00$ & $-1.00$ & $-1.00$ &   0.02\\
 \Sys{5} & 0.48 $\pm$ 0.09 & 0.53 $\pm$ 0.08 & 0.12 $\pm$ 0.05 & $+0.50$ & $+0.60$ & $-0.30$ &   0.25\\\hline
  \Sys{} & 0.59 $\pm$ 0.09 & 1.19 $\pm$ 0.06 & 0.56 $\pm$ 0.07 &         &         &         &   0.48\\
   Total & 0.69 $\pm$ 0.09 & 1.30 $\pm$ 0.06 & 0.61 $\pm$ 0.07 & $+0.29$ & $+0.29$ & $+0.35$ &   0.52\\\hline
   \chiq &                 &                 &                 &    3.00 &    4.28 &   0.25  &       \\
  Weight &            0.42 &           -0.00 &            0.59 &         &         &         &       \\
    Pull &            2.07 &           -1.08 &           -2.07 &         &         &         &       \\\hline
\end{tabular}
\end{center}
\caption{Combination of one observable \M\ from three correlated estimates \Mz,
  \Mo\ and \Mt\ using the \BLUE\ software. Counting from the left, columns 2-4
  show the individual results, their uncertainties and the statistical precision
  in the systematic uncertainties. Columns 5-7 report the estimator correlations
  \rhoijk\ for the pair of estimates~($i, j$) for all sources $k$ of
  uncertainty. The line denoted by Total lists the total uncertainties \si\ and
  correlations \rhoij.
  The rightmost column shows the combined result.
  The lower part of the table reports the estimator consistencies expressed as
  pairwise \chiq\ values, the weights of the estimates within the combination
  and the pulls of the estimates. Details on the calculation of \chiq, the
  weights and pulls are given in Ref.~\protect\cite{NIS-1401}.
\label{tab:exam}
}
\end{table}
%
 A compact example of three estimates of a single observable is listed in
 Table~\ref{tab:exam}. The code reproducing the content of this table and all
 following figures is given in~\ref{app:soft}.
 It contains detailed documentation relating the various function calls to the
 steps described in Fig.~\ref{fig:work}, and the obtained results to
 Table~\ref{tab:exam} and the respective figures.
 After installing the package, the only two commands to be executed from the
 shell prompt are:
%
 \begin{lstlisting}
 root -b < B_SoftExample.inp > B_SoftExample.out
 pdflatex  B_SoftExample.tex
\end{lstlisting}

 Although for this example the values chosen are similar to what is obtained in
 measurements of the top quark mass, this example is purely artificial.
 The estimate \Mz\ deliberately was chosen such as to have large \chiq\ values
 in the compatibility evaluations with the other two estimates.
 Since the three measurements should come from the same underlying \xT, this
 indicates either an unlikely outcome or a potential systematic problem with
 this result. Only after a careful investigation of this measurement, resulting
 in a low probability for the second possibility, should this measurement be
 included in the combination.
 The systematic uncertainties are shown together with the statistical precisions
 at which they are known. Those statistical precisions allow evaluating whether
 two estimators have a significantly different sensitivity for a source of
 uncertainty\footnote{Two quantities ($\xo\pm\so,\,\xt\pm\st$) are significantly
   different, if their difference ($\Delta=\xo-\xt$) with $\sdq = \soq + \stq -
   2\,\rhoot \so \st$ is significantly different from zero. In this case, the
   $x_i$ correspond to the systematic uncertainties and the \si\ to their
   statistical precisions.}. In addition, they indicate which systematic effect
 should be evaluated with higher statistical precision.
%
\begin{figure*}[tbp!]
\centering
\subfigure[\Mz\ vs.~\Mo]{\label{fig:cono}
  \includegraphics[width=0.48\textwidth]{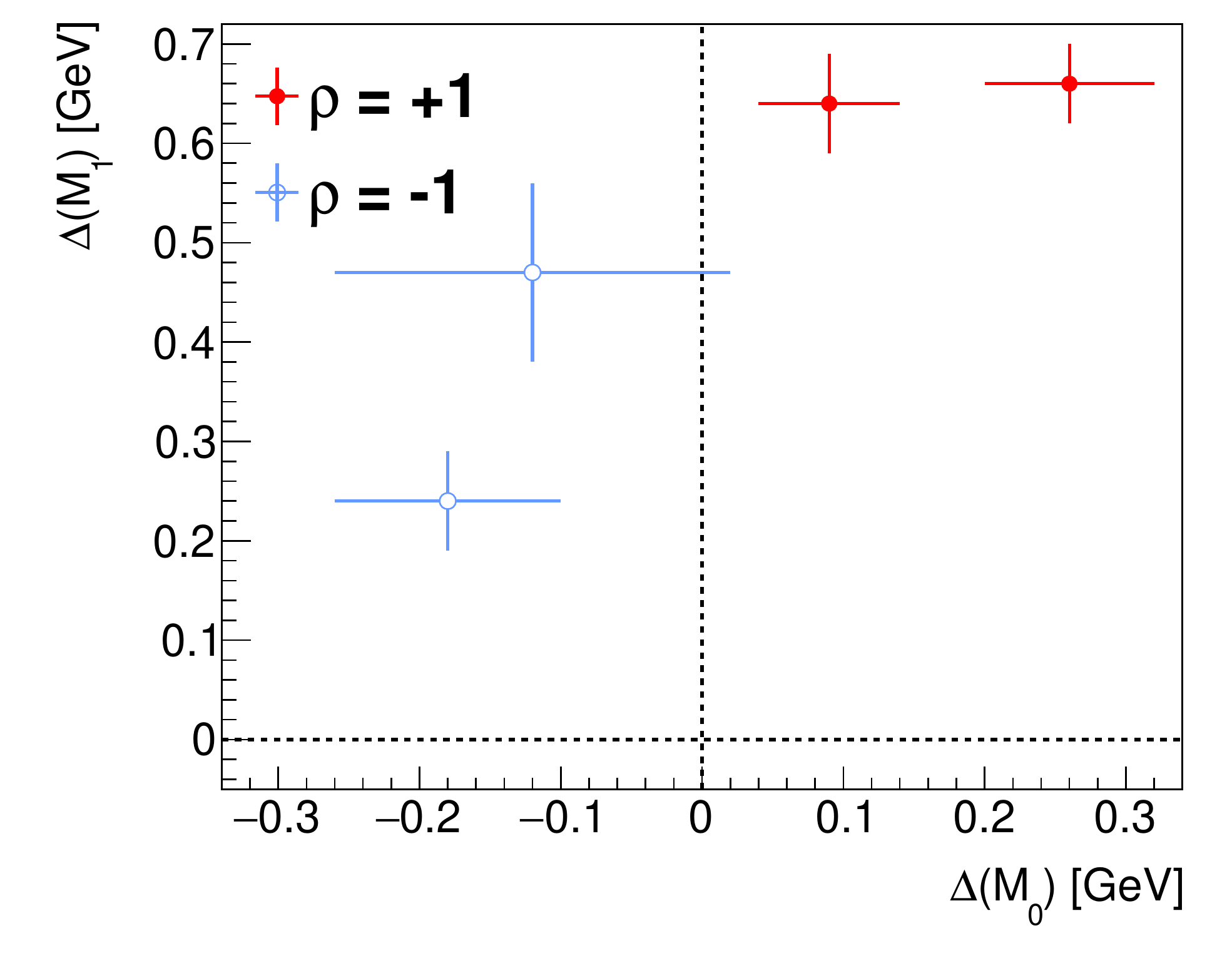}}
\subfigure[\Mz\ vs.~\Mt]{\label{fig:cont}
  \includegraphics[width=0.48\textwidth]{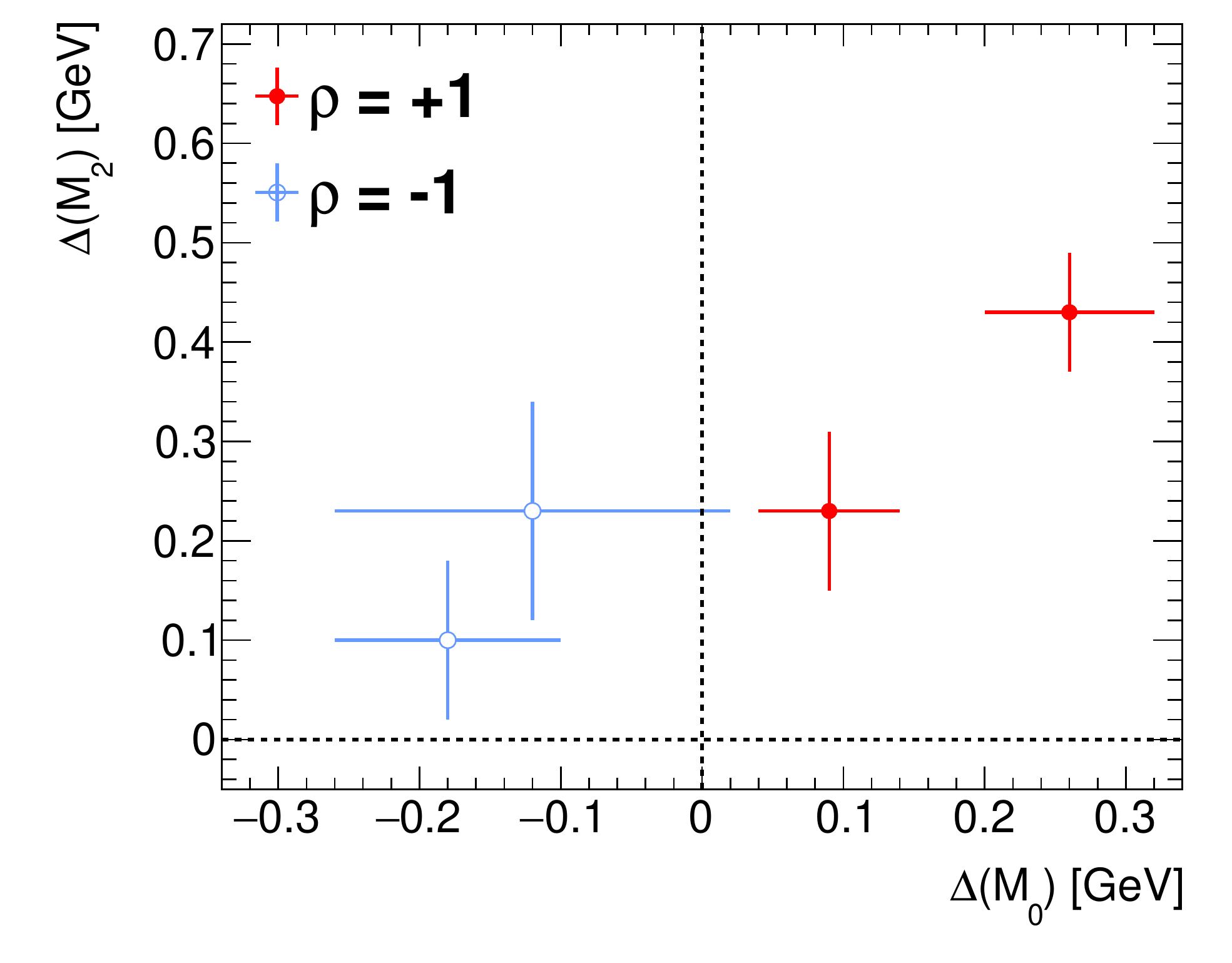}}
\caption{The sources of systematic uncertainties with estimator correlations of
  $\rhoijk=+1$~(full red points) or $\rhoijk=-1$~(open blue points). The points
  indicate the size of the uncertainties and the bars their statistical
  precision. In each of the figures, from right to left the points correspond to
  the sources \Sys{1} to \Sys{4} from Table~\protect\ref{tab:exam}.
\label{fig:corr}
}
\end{figure*}

 The sources of systematic uncertainties for which the estimator correlations
 are $\rhoijk=\pm1$ are shown in Fig.~\ref{fig:corr}. The case $\rhoijk=+1$
 corresponds to the situation where simultaneously applying a systematic effect
 to both estimates (e.g.~increasing a component $l$ of the jet energy scale
 uncertainty by $+1\sigJES$ as e.g.~performed in Ref.~\cite{TOPQ-2013-02}) leads
 to both measured values moving into the same direction, either both get larger
 or both get smaller than the original result. The case $\rhoijk=-1$ means the
 two measurements move in opposite directions. See Ref.~\cite{TOPQ-2013-02} for
 further details.
 The points for which the bars cross one of the coordinate axis indicate sources
 for which, within uncertainties, the correlation may be $\rhoijk=+1$ or
 $\rhoijk=-1$. For example, this is the case for \Sys{3} of \Mz\ from
 Table~\protect\ref{tab:exam}, i.e.~for the upper point in the left quadrant in
 both subfigures of Fig.~\ref{fig:corr}.
 This will be exploited in the stability evaluation discussed below.

\begin{figure*}[tbp!]
\centering
\subfigure[Combined value]{\label{fig:cova}
  \includegraphics[width=0.48\textwidth]{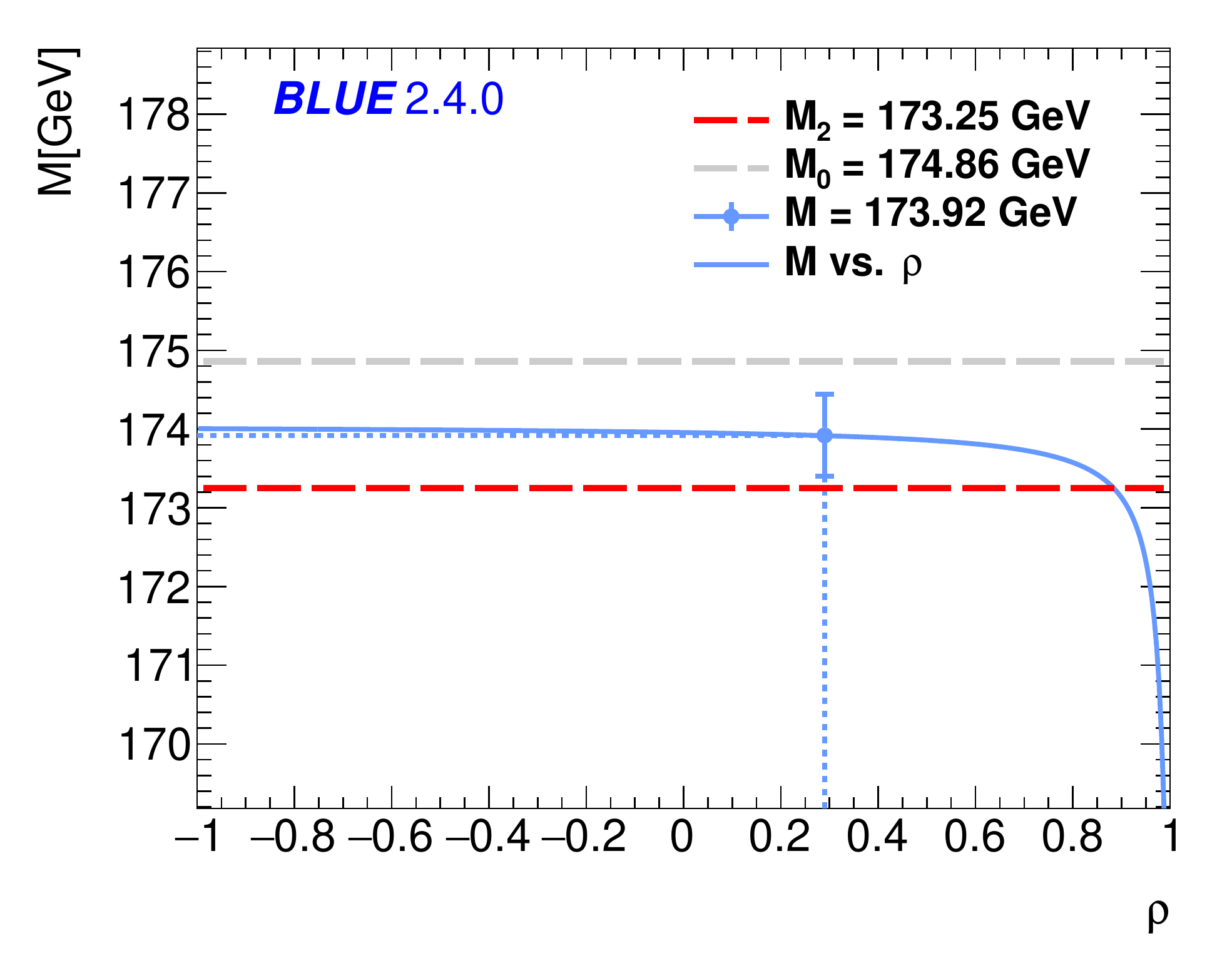}}
\subfigure[Uncertainty in the combined value]{\label{fig:coun}
  \includegraphics[width=0.48\textwidth]{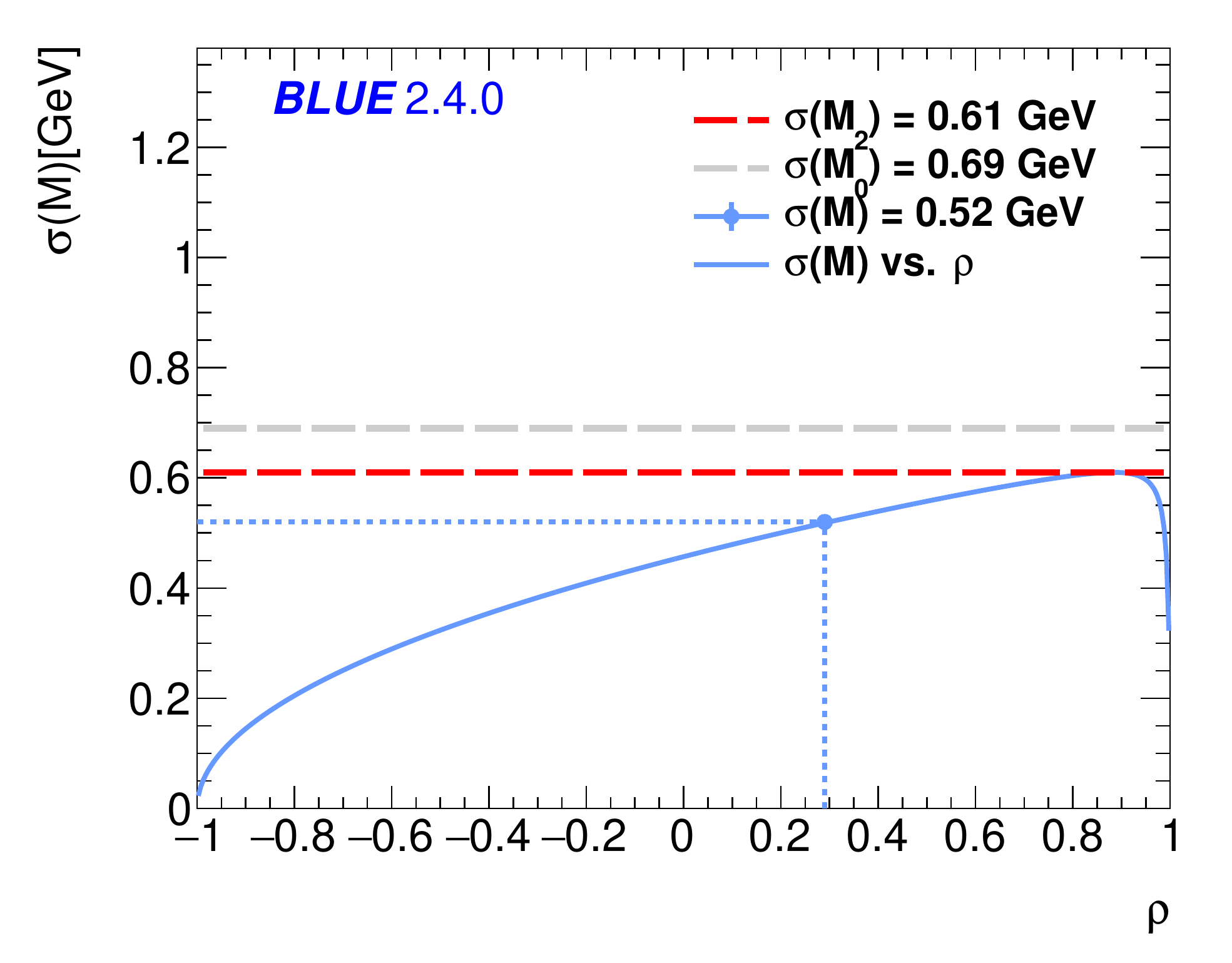}}
\caption{Results of the combination of the estimates \Mz\ and \Mt\ as functions
  of \rhof, where the blue point corresponds to the actual correlation
  $\rhof(\Mz, \Mt)$. The combined value is shown in~(a), the uncertainty in the
  combined value in~(b).
\label{fig:comb}
}
\end{figure*}

 Without combining, the precision of the knowledge about the observable is
 defined by the most precise result, here \Mt.
 The impact that an additional estimate has can be digested by performing
 pairwise combinations with the most precise result.
 An example of such a pairwise combination of \Mz\ and \Mt\ is shown in
 Fig.~\ref{fig:comb}.
 Apart from the range $\rhof>0.8$, the combined value is almost independent of
 \rhof. In contrast, the uncertainty in the combined value has a very strong
 dependence on \rhof.
%

\begin{figure*}[t!]
\centering
\subfigure[Full combination]{\label{fig:sola}
  \includegraphics[width=0.48\textwidth]{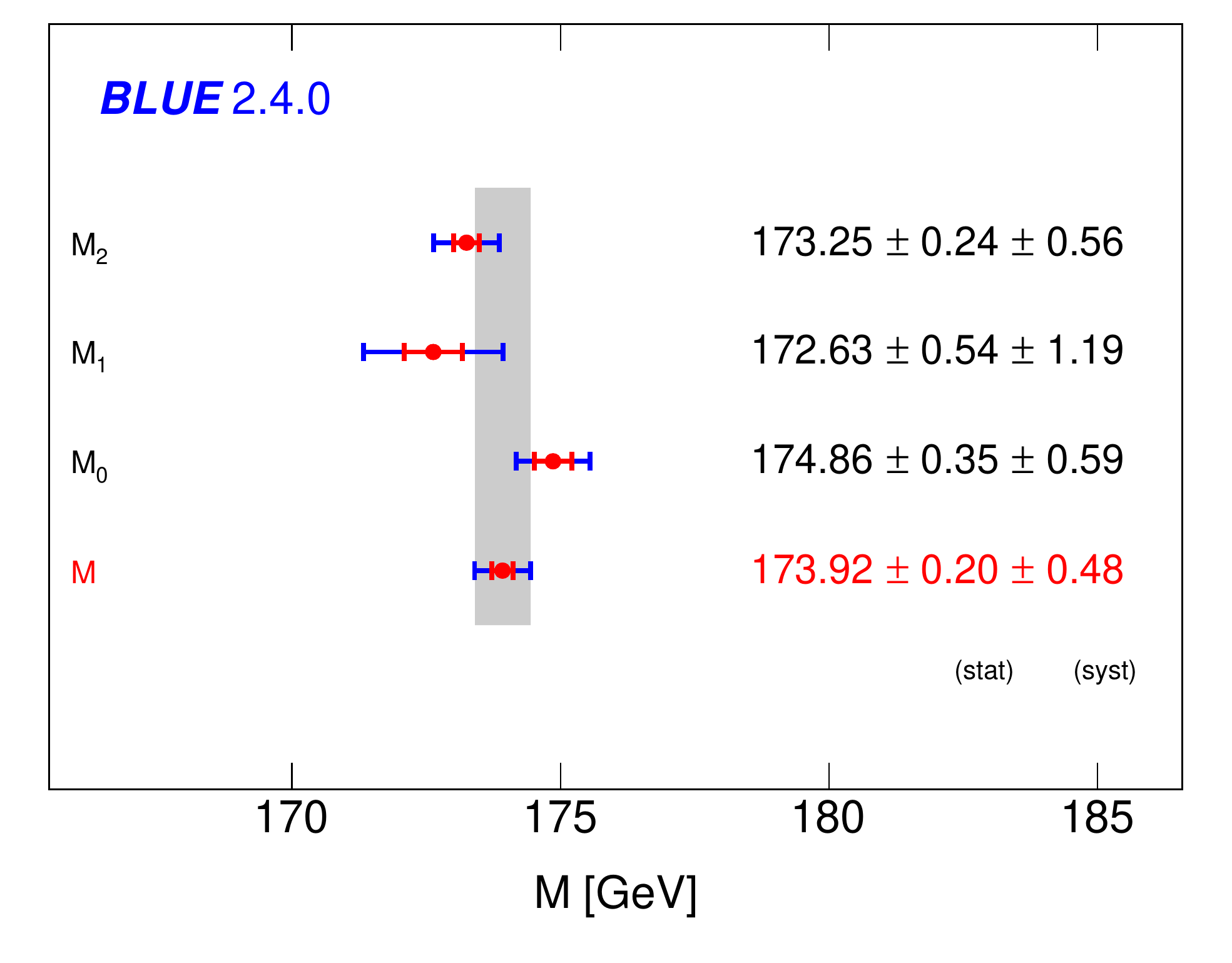}}
\subfigure[Successive combination]{\label{fig:sols}
  \includegraphics[width=0.48\textwidth]{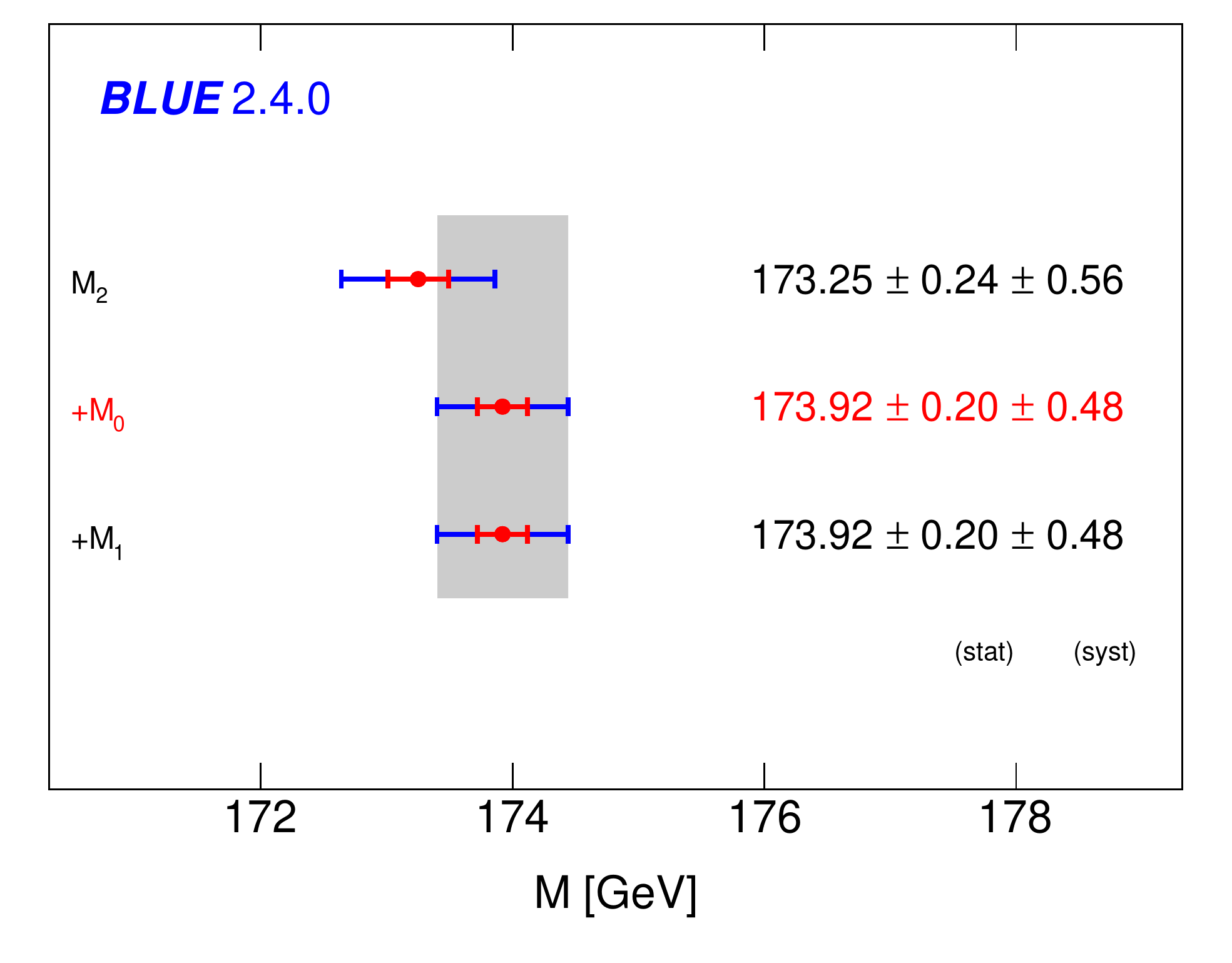}}
\caption{The three estimates and the combined value of all estimates are shown
  in~(a). In contrast, (b) shows the combined results obtained in successive
  combinations adding one estimate at a time, i.e.~the second line is the result
  of combining \Mt and \Mz, the third line the result of the combination of all
  three measurements.
\label{fig:solv}
}
\end{figure*}

 The combination of all estimates is shown in Fig.~\ref{fig:sola}. The input
 measurements are listed in the first three lines, the combined result is listed
 in red in the last line.
 Fig.~\ref{fig:sols} reveals that not all results significantly contribute to
 the combined value. In this figure, the lines show the results of successive
 combinations, always adding the estimate listed, to the previous list of
 estimates.
 Also here, the suggested combined result is shown in red. At the quoted
 precision, the estimate \Mo\ does not improve the already accumulated result
 obtained from combining \Mt\ and \Mz. This means \Mo\ merely serves as a
 cross-check measurement for the combination.

\begin{figure*}[tbp!]
\centering
\subfigure[Combined value]{\label{fig:vast}
  \includegraphics[width=0.48\textwidth]{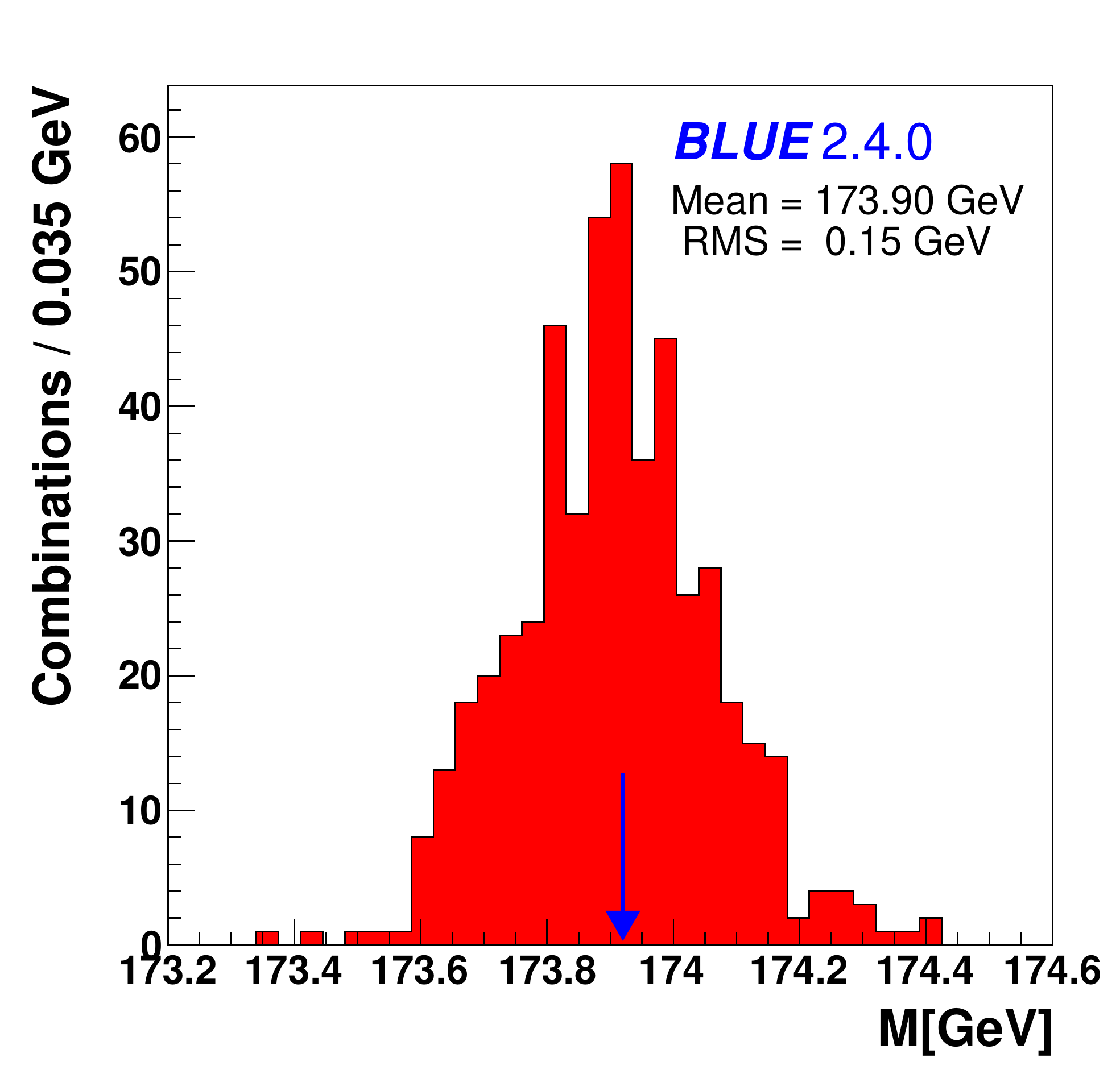}}
\subfigure[Uncertainty in combined value]{\label{fig:unst}
  \includegraphics[width=0.48\textwidth]{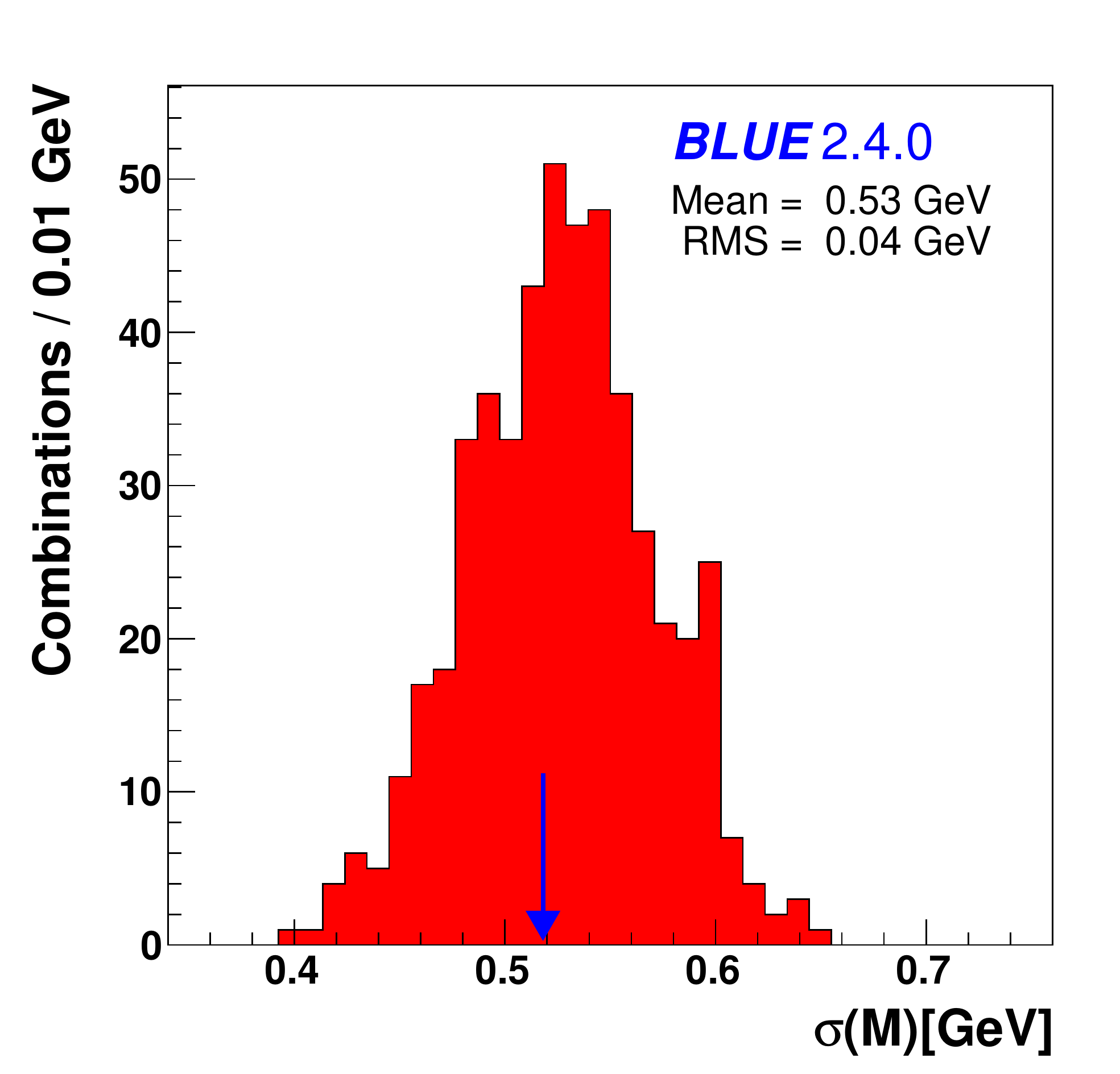}}
\caption{The stability of the combined value~(a) and the uncertainty in the
  combined value~(b), obtained in 500 combinations, while varying the input
  uncertainties and correlations according to their statistical precision.
\label{fig:stab}
}
\end{figure*}

 Fig.~\ref{fig:stab} shows the stability of the combination of all three
 results, taking into account the statistical precisions at which the systematic
 uncertainties are known, see Table~\ref{tab:exam}.
 For this figure, all systematic uncertainties are altered within their
 statistical precisions. For sources with $\rhoijk=\pm 1$ also the correlations
 are re-evaluated, i.e.~they may change sign, see Fig.~\ref{fig:corr}.
 The varied input measurements are combined. The resulting combined values and
 uncertainties in the combined values are shown in the histograms.
 For the statistical precision at which the uncertainties are known, the
 combined result is uncertain by $0.15$~GeV and the related uncertainty by
 $0.04$~GeV.
 These uncertainties exist on top of what is quoted in Table~\ref{tab:exam}.
 Frequently those are not provided, or even not evaluated. This is only
 justified if they are much smaller than the quoted uncertainties.
 For the statistical uncertainty in the combined result of $0.20$~GeV, see
 Table~\ref{tab:exam}, the situation is at the border of being acceptable.
%
\section{Impact}
\label{sec:impa}
%
%
 The software can be used for an in-depth analysis of the impact of various
 assumptions made in the combination. In case the relevant input is provided, it
 also allows assessing the stability of the combination.

 Because of the large reduction in the uncertainty in the combined result
 obtained by lowering the estimator correlations, see Fig.~\ref{fig:comb}, it
 is advisable to use this software already in the design stage of the various
 analyses performed for obtaining the same observable within a single
 experiment.
 Usually, the uncertainties in the various systematic effects (e.g.~the
 uncertainty in jet energy scales for experiments at hadron collider) are
 determined by the actual level of understanding of the detector and have to be
 taken into account at face value. In contrast, the sensitivity of the
 estimators to those effects can be influenced by the estimator design.
 This way their correlation can be reduced, thereby improving the gain obtained
 in the combination. Generally speaking, the strategy should not be to take over
 an aspect of the analysis that has worked for one estimator to another
 estimator. Instead, alternative approaches should be pursuit, such as to
 potentially lower the estimator correlations, even at the expense of a larger
 uncertainty.
 This is because achieving an anti-correlated pair of estimates with the same
 sensitivity to a specific source of uncertainty, renders this a significantly
 smaller uncertainty in the combined result.
 This can be seen for the sources \Sys{3}\ and \Sys{4}, for which
 $\rho_{ij3}=\rho_{ij4}=-1$. Those sources exhibit the largest fractional gain
 in uncertainty when comparing \M\ with \Mz, e.g. $\sigma_{x4}/\sigma_{24} =
 0.02/0.10 = 1/5$.
%
%
 An example of such an optimization is explained in Ref.~\cite{TOPQ-2013-02}.
 This software can be of significant help in this process.
 
 According to the knowledge of the author, by now the \BLUE\ software was used
 in a number of combinations, mostly in the context of high energy physics,
 especially at the Large Hadron Collider~(LHC).
 Examples from the ALICE, ATLAS, CMS and LHCb collaborations are detailed in
 Refs.~\cite{ALIC-2018-01,TOPQ-2017-03,CMS-2017-01,LHCB-2018-01}.
 The first world combination of the top quark mass~\cite{CONF-2014-008} has also
 been performed with this software.
 In addition to the LHC collaborations, the software has been used by the
 PHENIX~\cite{PHENIX-2018-01} and STAR~\cite{STAR-2016-01} collaborations, and
 in a combination of the strong coupling constant \as\ from many results in
 Ref.~\cite{ENT-2019-01}.
 Further examples of the software usage are described in the manual listed in
 the Code metadata table.
 To assist the users in developing their own combination code, the corresponding
 \verb!C++! routines to reproduce those published results are included in the
 software package.
 Although the above examples are all particle physics applications, the use of
 this software is not confined to a specific area of research. Any set of
 correlated measurements of one or more observables can be combined.
%
\section{Conclusions}
\label{sec:conc}
 The software performs the combination of $m$ correlated estimates of $n$
 physics observables~($m\ge n$) using the Best Linear Unbiased Estimate (\BLUE)
 method.
 The large flexibility, together with the several implemented correlation models
 and combination methods makes it a useful tool to assess details on the
 combination in question.
 Exploring the combination of various estimators of the same observable within a
 single experiment allows a design of estimators with low correlation.
 This enhances the gain achieved in combinations of estimates obtained from
 those estimators.
%
\section*{Declaration of competing interest}
 The author declares that he has no known competing financial interests or
 personal relationships that could have appeared to influence the work reported
 in this paper.
%
\clearpage
\appendix
\section{Example code}
\label{app:soft}
\footnotesize
\begin{lstlisting}[language=C++,frame=left,numbers=left,%
  caption={B\_SoftExample.inp}]
// Load Blue library
gSystem->Load("libBlue.so");

// Compile and execute code. This will produce
// the below '.cxx' files and a '.tex' file
.L B_SoftExample.cxx++
   B_SoftExample()

// Figure 3a
.L B_SoftExample_x0_x1_CorPai.cxx++
   B_SoftExample_x0_x1_CorPai()

// Figure 3b
.L B_SoftExample_x0_x2_CorPai.cxx++
   B_SoftExample_x0_x2_CorPai()

// Figure 4a,b
.L B_SoftExample_x0_x2_DisPai.cxx++
   B_SoftExample_x0_x2_DisPai()

// Figure 5a
.L B_SoftExample_DisRes_Obs_0.cxx++
   B_SoftExample_DisRes_Obs_0()

// Figure 5b
.L B_SoftExample_AccImp_Obs_0.cxx++
   B_SoftExample_AccImp_Obs_0()
.q
\end{lstlisting}
\clearpage
\begin{lstlisting}[language=C++,frame=left,numbers=left,%
  caption={B\_SoftExample.cxx}]
#include "Blue.h"

void B_SoftExample(){

  //-- Start preparing the inputs for later use in BLUE
  // Numbers of estimates, uncertainties and observables
  static const Int_t NumEst = 3;
  static const Int_t NumUnc = 6;
  static const Int_t NumObs = 1;
  
  // Array of names of estimates (i,j = 0, 1, 2)
  static const TString NamEst[NumEst] = {"M_{0}", "M_{1}", "M_{2}"};

  // Array of names of uncertainties (k = 0, ..., 5)
  static const TString NamUnc[NumUnc] = {
    "Stat", "Syst_{1}", "Syst_{2}", "Syst_{3}", "Syst_{4}", "Syst_{5}"};

  // Array of names of observables (n = 0)
  static const TString NamObs[NumObs] = {"M"};

  // Array of estimates and uncertainties, copy to matrix
  static const Int_t LenXEst = NumEst * (NumUnc+1);
  static const Double_t XEst[LenXEst] = {
    //      Stat  Sys1  Sys2  Sys3  Sys4  Sys5
    174.86, 0.35, 0.26, 0.09, 0.12, 0.18, 0.48,
    172.63, 0.54, 0.66, 0.64, 0.47, 0.24, 0.53,
    173.25, 0.24, 0.43, 0.23, 0.23, 0.10, 0.12};
  TMatrixD* InpEst = new TMatrixD(NumEst, NumUnc+1, &XEst[0]);
  
  // Array of statistical precision in systematic uncertainties, copy to matrix
  static const Int_t LenSUnc = NumEst * NumUnc;
  static const Double_t SUnc[LenSUnc] = {
    //Stat  Sys1  Sys2  Sys3  Sys4  Sys5
    0.00,   0.06, 0.05, 0.14, 0.08, 0.09,
    0.00,   0.04, 0.05, 0.09, 0.05, 0.08,
    0.00,   0.06, 0.08, 0.11, 0.08, 0.05};
  TMatrixD* InpSta = new TMatrixD(NumEst, NumUnc, &SUnc[0]);

  // Array of rho_ijk == const correlations for k < 5
  static const Double_t RhoVal[NumUnc-1] = {
    0.0, 1.0, 1.0, -1.0, -1.0};

  // Array for correlation matrix for k == 5 with rho_ij5 != const
  static const Int_t LenInd = NumEst * (NumEst-1) / 2;
  static const Double_t RhoInd[LenInd] = {0.5, 0.6, -0.3};

  // Formats and file name for figures and Latex file
  static const TString ForVal = "%5.2f";
  static const TString ForUnc = "%4.2f";
  static const TString ForWei = ForUnc;
  static const TString ForRho = "%+4.2f";
  static const TString ForPul = ForUnc;
  static const TString ForChi = ForUnc;
  static const TString ForUni = "GeV";
  static const TString FilBas = "B_SoftExample";

  // Axis ranges for Figs.3ab and Fig.6ab
  static const Double_t XvaMax = 0.34, YvaMin = -0.05, YvaMax = 0.72;
  static const Double_t ValMin = 173.2, ValMax = 174.6;
  static const Double_t UncMin =  0.34, UncMax =  0.76;
  //-- End of preparation of inputs for BLUE

  //-- Start using BLUE --- no inputs below this line ---
  //-- Keywords: relate to the flowchart of the software
  // Create: call constructor of BLUE object
  Blue *myBlue = new Blue(NumEst, NumUnc);

  // Fill: set display formats and define names
  myBlue->SetFormat(ForVal, ForUnc, ForWei, ForRho, ForPul, ForChi, ForUni);
  myBlue->FillNamEst(&NamEst[0]);
  myBlue->FillNamUnc(&NamUnc[0]);
  myBlue->FillNamObs(&NamObs[0]);
  
  // Fill: insert estimates with their uncertainties
  myBlue->FillEst(InpEst);
       
  // Fill: insert statistical precision in systematic uncertainties
  myBlue->FillSta(InpSta);

  // Fill: insert the estimator correlations for all sources of uncertainty
  for(Int_t k = 0; k<NumUnc; k++){
    if(k != 5){myBlue->FillCor(k, RhoVal[k]);
    }else{myBlue->FillCor(-k, &RhoInd[0]);}
  }

  // Change:
  // ... alterations would go here

  // Fix:
  myBlue->FixInp();
  
  // Combine:
  myBlue->Solve();

  // Digest: show some estimate quantities, prepare code for Figs. 3 and 4
  myBlue->PrintEst();
  myBlue->CorrelPair(0, 1, FilBas, -XvaMax, XvaMax, YvaMin, YvaMax);
  myBlue->CorrelPair(0, 2, FilBas, -XvaMax, XvaMax, YvaMin, YvaMax);
  myBlue->DisplayPair(0, 2, FilBas);

  // Digest: show some observable quantities, prepare code for Fig. 5a
  myBlue->PrintChiPro();
  myBlue->DisplayResult(0, FilBas);

  // Digest: write Latex file
  myBlue->LatexResult(FilBas);

  // Free:, Fix:, Combine: solve according to importance
  myBlue->ReleaseInp();
  myBlue->FixInp();
  myBlue->SolveAccImp(0.1);

  // Digest: prepare code for Fig. 5b
  myBlue->DisplayAccImp(0, FilBas);

  // Free:, Fix:, Combine: solve varying the systematic uncertainties
  myBlue->ReleaseInp();
  myBlue->FixInp();
  myBlue->SolveScaSta();

  // Digest: produce Fig. 6
  myBlue->PrintScaSta(FilBas, ValMin, ValMax, UncMin, UncMax);

  // Delete: delete object as well as local matrices and return
  delete myBlue; myBlue = NULL;  
  InpEst->Delete(); InpEst = NULL; 
  InpSta->Delete(); InpSta = NULL; 
  return;
};

\end{lstlisting}
%
%
\normalsize
\bibliographystyle{elsarticle-num} 
\bibliography{BlueSoft.bib}
\end{document}